\documentclass[sigconf]{acmart}
\usepackage{graphicx}
\usepackage{ulem}
\usepackage{verbatim}
\usepackage{multirow}
\usepackage[linesnumbered, ruled, vlined]{algorithm2e}
\usepackage{longtable}
\usepackage{array}
\usepackage{subfigure}
\usepackage{stfloats}
\usepackage{balance}
\usepackage{multicol}
\usepackage{color}
\usepackage{bm}
\usepackage{booktabs} 
\usepackage{epsfig}
\usepackage{enumitem}
\usepackage{balance}

\usepackage{arydshln}

\usepackage{cleveref}
\usepackage{graphicx}
\usepackage{textcomp}
\usepackage{xcolor}
\usepackage{subfigure}

\usepackage{tabularx}

\newcommand{\hide}[1]{} 
\newcommand{\vpara}[1]{\vspace{0.1in}\noindent\textbf{#1}}

\newcommand{\secref}[1]{Section~\ref{#1}} 
\newcommand{\beq}[1]{\vspace{-0.02in}\begin{equation}#1\end{equation}\vspace{-0.02in}}
\newcommand{\beqn}[1]{\vspace{-0.03in}\begin{eqnarray}#1\end{eqnarray}\vspace{-0.03in}}

\newcommand{\RC}{MrGCN}
\newcommand{\sRC}{MrGCN\space}

\AtBeginDocument{%
  \providecommand\BibTeX{{%
    \normalfont B\kern-0.5em{\scshape i\kern-0.25em b}\kern-0.8em\TeX}}}

\setcopyright{acmcopyright}
\copyrightyear{2021}
\acmYear{2021}
\acmDOI{10.1145/1122445.1122456}

\acmConference[WSDM '21]{The 14th ACM International Conference on Web Search and Data Mining}{March 08--12, 2021}{Jerusalem, Israel}
\def\acmBooktitle#1{\gdef\@acmBooktitle{#1}}
\acmBooktitle{WSDM '21: The 14th ACM International Conference on Web Search and Data Mining,March 08--12, 2021, WSDM, Jerusalem}
\acmPrice{15.00}
\acmISBN{978-1-4503-XXXX-X/18/06}

\begin{document}

\title{Pre-Training Graph Neural Networks for Cold-Start Users and Items Representation}
\renewcommand{\shorttitle}{Pre-Training GNNs for Cold-Start Users and Items Representation}


\author{Bowen Hao}
\affiliation{
	\institution{Renmin University of China}
}
\email{ jeremyhao@ruc.edu.cn    }

\author{Jing Zhang}
\authornote{Corresponding Author.}
\affiliation{
	\institution{Renmin University of China}
}
\email{ zhang-jing@ruc.edu.cn    }

\author{Hongzhi Yin}
\affiliation{%
	\institution{The University of Queensland}
}
\email{h.yin1@uq.edu.au}

\author{Cuiping Li}
\affiliation{
	\institution{Renmin University of China}
}
\email{licuiping@ruc.edu.cn}

\author{ Hong Chen}
\affiliation{
	\institution{Renmin University of China}
}
\email{chong@ruc.edu.cn}


\renewcommand{\shortauthors}{B.Hao et al.}
\begin{abstract}
Cold-start problem is a fundamental challenge for recommendation tasks. 
Despite the recent advances on Graph Neural Networks (GNNs) incorporate the high-order collaborative signal to alleviate the problem, the embeddings of the cold-start users and items aren't explicitly optimized, and the cold-start neighbors are not dealt with during the graph convolution in GNNs.
This paper proposes to pre-train a GNN model before applying it for recommendation. 
Unlike the goal of recommendation, the pre-training GNN simulates the cold-start scenarios from the users/items with sufficient  interactions and takes the embedding reconstruction as the pretext task, such that it can directly improve the embedding quality and can be easily adapted to the new cold-start users/items. 
To further reduce the impact from the cold-start neighbors, we incorporate a self-attention-based meta aggregator to enhance the aggregation ability of each graph convolution step, and an adaptive neighbor sampler to select the effective neighbors according to the feedbacks from the pre-training GNN model. Experiments on three public recommendation datasets show the superiority of our pre-training GNN model against the original GNN models on user/item embedding inference and the recommendation task.

\end{abstract}

\ccsdesc[500]{Information systems~Social advertising}

\keywords{Pre-training, graph neural networks, cold-start, recommendation}


\maketitle

\begin{figure}[t]
	\centering
	\includegraphics[width= 0.45 \textwidth]{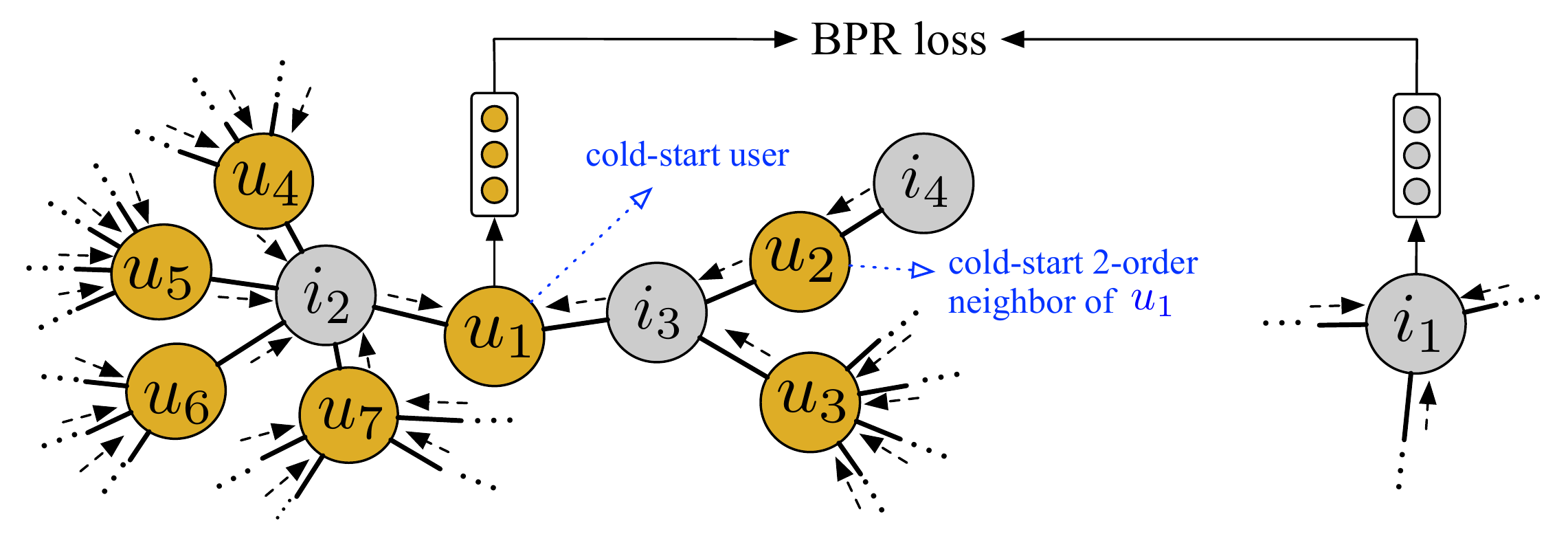}
	\caption{\label{fig:GNN_recommendation} A GNN model for recommendation. }
\end{figure}

\section{Introduction}

Recommendation systems~\cite{he2017neural,linden2003amazon} have been extensively deployed to alleviate information overload in various web services, such as social media, E-commerce websites and news portals. To predict the likelihood of a user adopting an item, collaborative filtering (CF) is the most widely adopted principle. The most common paradigm  for CF, such as matrix factorization~\cite{linden2003amazon} and neural collaborative filtering~\cite{he2017neural}, is to learn embeddings, i.e. the preferences for users and items and then perform the prediction based on the embeddings~\cite{xiangnanhe_lightgcn20}. However, these models fail to learn high-quality embeddings for the cold-start users/items with sparse interactions. 


To address the cold-start problem, traditional recommender systems  incorporate the side information such as content features of users and items~\cite{hongzhi_side,YinWWCZ17} or external knowledge graphs (KGs)~\cite{wang2019multi,wang2019kgat} to compensate the low-quality embeddings caused by sparse interactions. 
However, the content features are not always available, and it is not easy to link the items to the entities in KGs due to the incompleteness and ambiguation of the entities. 

On another line, inspired by the recent development of  graph neural networks (GNNs)~\cite{Thomasgcn,williamgraphsage17,hongxugraphcsc19}, NGCF~\cite{wangncgf19} and LightGCN~\cite{xiangnanhe_lightgcn20} encode the high-order collaborative signal in the user-item interaction graph by a GNN model, based on which they perform the recommendation task.
As shown in Fig.~\ref{fig:GNN_recommendation}, a typical recommendation-oriented GNN conducts graph convolution on the local neighborhood’s  embeddings of $u_1$ and $i_1$. Through iteratively repeating the convolution by multiple steps, the embeddings of the high-order neighbors are propagated to $u_1$ and $i_1$. Based on the aggregated embeddings of $u_1$ and $i_1$, the likelihood of $u_1$ adopting $i_1$ is estimated, and cross-entropy loss~\cite{chenhongxutked20} or BPR loss~\cite{wangncgf19,xiangnanhe_lightgcn20} is usually adopted to compare the likelihood and the true observations.


Despite the success of capturing the high-order collaborative signal in GNNs~\cite{wangncgf19,xiangnanhe_lightgcn20}, the cold-start problem is not thoroughly solved by them. First, the GNNs for recommendation address the cold-start user/item embeddings through optimizing the likelihood of a user adopting an item, which isn't a direct improvement of the embedding quality; second, the GNN model does not specially deal with the cold-start neighbors among all the neighbors when performing the graph convolution.
For example in Fig.~\ref{fig:GNN_recommendation}, to represent $u_1$,  the 2-order neighbor $u_2$ is also a cold-start user who only interacts with $i_3$ and $i_4$. The result of graph convolution on the inaccurate embedding of $u_2$ and the embedding of $u_3$ together will be propagated to $u_1$ and hurt its embedding.
Existing GNNs ignore the cold-start characteristics of neighbors during the graph convolution process. Although some GNN models such as GrageSAGE~\cite{williamgraphsage17} or FastGCN~\cite{chenfastgcn18} filter neighbors before aggregating them, they usually follow a random or an importance sampling strategy, which also ignore the cold-start characteristics of the neighbors. This leads us to the following research problem: \textit{how can we learn more accurate embeddings for cold-start users or items by GNNs?}

\vpara{Present work.} 
To tackle the above challenges, before performing the GNN model for recommendation, we propose to pre-train the GNN model to enhance the embeddings of the cold-start users or items. 
Unlike the goal of recommendation, the pre-training task directly reconstructs the cold-start user/item embeddings by mimicking the meta-learning setting via episode based training, as proposed in~\cite{vinyalsmatching16}.
Specifically, we pick the users/items with sufficient interactions as the target users/items and learn their ground truth embeddings on the observed abundant interactions. 
To simulate the real cold-start scenarios, in each training episode, we randomly sample $K$ neighbors for each target user/item, based on which we perform the graph convolution multiple steps to predict the target embedding. The reconstruction loss between the predicted embedding and the ground truth embedding is optimized to directly improve the embedding capacity, making the model easily and rapidly being adapted to new cold-start users/items.


However, the above pre-training strategy still can not explicitly deal with the high-order cold-start neighbors when performing graph convolution.
Besides, previous GNN sampling strategies such as random or importance sampling strategies may fail to sample high-order relevant cold-start neighbors due to their sparse interactions.
To overcome these challenges, we incorporate  a meta aggregator and an adaptive neighbor sampler into the pre-training GNN model. Specifically, the meta aggregator learns cold-start users/items' embeddings on the first-order neighbors by self-attention mechanism under the same meta-learning setting, which is then incorporated into each graph convolution step to enhance the aggregation ability. 
While the adaptive neighbor sampler is formalized as a hierarchical Markov Sequential Decision Process, which sequentially samples from the low-order neighbors to the high-order neighbors according to the feedbacks provided by the pre-training GNN model.
The two components are jointly trained. Since the GNN model can be instantiated by different choices such as the original GCN~\cite{Thomasgcn}, GAT~\cite{gat18} or FastGCN~\cite{chenfastgcn18}, the proposed pre-training GNN model is model-agnostic. The contributions of this work are as follows:


\begin{itemize}[ leftmargin=10pt ]
	\item We propose a pre-training GNN model to learn high-quality embeddings for cold-start users/items. The model is learned under the meta-learning setting to reconstruct the user/item embeddings, which has the powerful generalization capacity.   
	\item To deal with the cold-start neighbors during the graph convolution process, we further propose a meta aggregator to enhance the aggregation ability of each graph convolution step, and a neighbor sampler to select the effective neighbors adaptively according to the feedbacks of the pre-training GNN model. 
	\item Experiments on both intrinsic embedding evaluation task and extrinsic downstream recommendation task demonstrate the superiority of our proposed pre-training GNN model against the state-of-the-art GNN models.
	
	\hide{Compared with the original GNN models, the pre-training GNN models achieve 15.3-53.4\% improvement in terms of Spearman correlation~\cite{ziniuhufewshot19} between the predicted embeddings and the ground truth embeddings by the intrinsic evaluation on inferring user/item embeddings, and also achieve  6.9-13.2\% improvement in NDCG@20 by the extrinsic evaluation on the downstream recommendation task.}
	
\end{itemize}

\begin{table}[t]
	
\end{table}

\section{Preliminaries}

In this section, we first define the problem and then introduce the graph neural networks that can be used to solve the problem.

We formalize the user-item interaction data for recommendation as a bipartite graph denoted as $G=(U,I,E)$, where $U = \{  u_1, \cdots, u_{|U|}  \}$ is the set of users and $I = \{  i_1, \cdots, i_{|I|} \}$ is the set of items. $U$ and $I$ comprise two types of the nodes in $G$. Notation $E \subseteq U \times I$ denotes the set of edges that connect the users and items. 

We use $\mathcal{N}^{l}(u)$ to represent the $l$-order neighbors of user $u$. When ignoring the superscript,  $\mathcal{N}(u)$ indicates the first-order neighbors of $u$. Similarly, $\mathcal{N}^{l}(i)$ and $\mathcal{N}(i)$ are defined for items.

Let $f:U \cup V \rightarrow \mathbb{R}^d$ be the encoding function that maps the users/items to $d$-dimension real-valued vectors. We use $\textbf{h}_{u}$ and $\textbf{h}_{i}$ to denote the embedding of user $u$ and item $i$ respectively.  
Given a bipartite graph $G$, we aim to pre-train the encoding function $f$ that is able to be applied on the downstream recommendation task to improve its performance. In the following sections, we mainly take user embedding as an example to explain the proposed model. Item embedding can be explained in the same way. 
	
\subsection{GNN for Recommendation}
The encoding function $f$ can be instantiated by various GNNs. Take GraphSAGE as an example, we first sample neighbors for each user $u$  randomly and then perform the graph convolution

\beqn{
	\label{eq:GraphSage}
\textbf{h}^l_{\mathcal{N}(u)} &=& \text{AGGREGATE}(\{\textbf{h}^{l-1}_i, \forall i \in \mathcal{N}(u)\}),	\\ \nonumber
\textbf{h}_{u}^l &=& \sigma (\mathbf{W}^l \cdot {\rm CONCAT} ( \textbf{h}_{u}^{l-1}  ,   \textbf{h}^{l}_{  \mathcal{N}(u)}  ), 
}

\noindent to obtain the embedding of $u$, where $l$ denotes the current convolution step and $\textbf{h}^l_{u}$ denotes user $u$'s embedding at this step. Similarly, we can obtain the item embedding $\textbf{h}^l_{i}$ at the $l$-th convolution step. Once the embeddings of the last step $L$ for all the users and items are obtained,  we calculate the relevance score $y(u, i) = {\textbf{h}_{u}^L}^\mathrm{T} \textbf{h}_{i}^L$ between user $u$ and item $i$ and adopt the BPR loss~\cite{xiangnanhe_lightgcn20, wangncgf19}, i.e., 

\beqn{
	\label{bprloss}
	\mathcal{L}_{BPR} = \sum_{ (u,i)\in E, (u,j)\notin E }  - \ln \sigma( y(u, i) - y(u, j) ) + \lambda || \Theta_{gnn} ||_2^2,
}

\noindent 
to optimize the user preferences over items. 

The above presented GNNs are end-to-end models that can learn user/item embeddings and then recommend items to users simultaneously.  For addressing the cold-start users/items, the GNNs can incorporate the high-order collaborative signal through iteratively repeating the sampling and the convolution processes. However, the goal of recommendation shown in Eq.\eqref{bprloss} can not explicitly improve the embedding quality of the cold-start users/items.

\section{The Pre-training GNN Model}
\begin{figure*}[t]
	\centering
	\includegraphics[width= \textwidth]{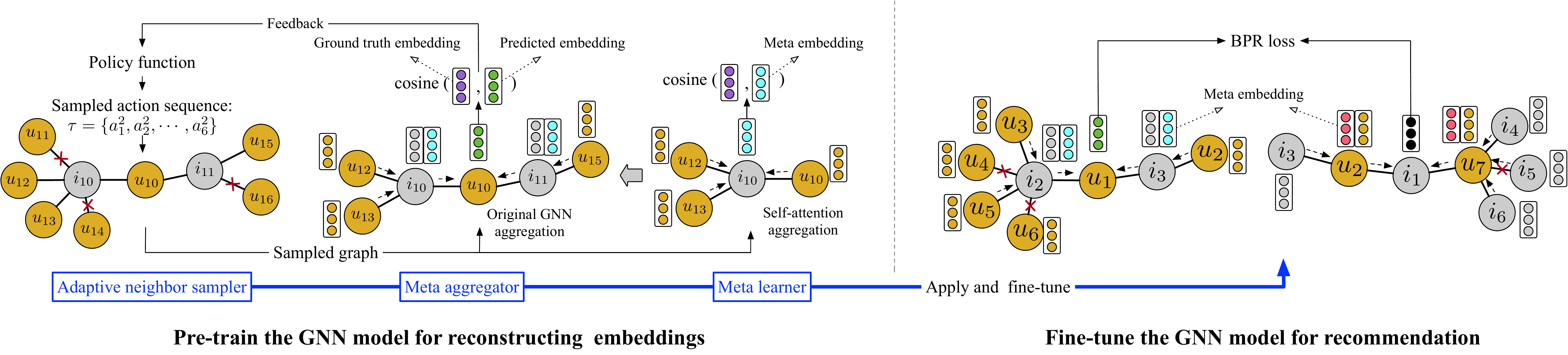}
	\caption{\label{fig:preraining-GNN} The overall framework of pre-training and fine-tuning the GNN model for recommendation. \small{The pre-training GNN model contains a meta aggregator which has incorporated a self-attention-based meta learner at each step of the original GNN aggregation, and a neighbor sampler which samples the neighbors adaptively according to the feedbacks from the cosine similarity between the predicted embedding and the ground truth embedding. The pre-trained GNN model is applied and fine-tuned on the downstream recommendation task. 
	} }
\end{figure*}

This section introduces the proposed pre-training GNN model to learn the embeddings for the cold-start users and items. We first describe a basic pre-training GNN model, and then explain a meta aggregator and an adaptive neighbor sampler that are incorporated in the model to further improve the embedding performance. Finally we explain how the model is fine-tuned on the downstream recommendation task.
The overview framework is shown in Fig.~\ref{fig:preraining-GNN}. 

\subsection{The Basic Pre-training GNN Model}
\label{sec:basic_pretraining_gnn}
We propose a basic pre-training GNN model to reconstruct the cold-start user/item embeddings in the meta-learning  setting. 
To achieve the goal, we need abundant cold-start users/items as the training instances. Since we also need ground truth embeddings of the cold-start users/items to learn $f$, we simulate those users/items from the target users/items with abundant interactions. The ground truth embedding for each user $u$, i.e., $\textbf{h}_u$, is learned upon the observed abundant interactions by NCF\footnote{The matrix factorization-based model is good enough to learn high-quality user/item embeddings from the abundant interactions.}~\cite{he2017neural}.
To mimic the cold-start users/items, in each training episode, we randomly sample $K$ neighbors for each target user/item. We repeat the sampling process $L$-1 steps from the target user to the $L$-1-order neighbors, which results in at most $K^l (1 \leq l \leq L)$ $l$-order neighbors for each target user/item. 
Similar to GraphSAGE~\cite{williamgraphsage17}, we sample high-order neighbors to improve the computational efficiency.  
Upon the  sampled first/high-order neighbors for the target user $u$, the graph convolution described in Eq.~\eqref{eq:GraphSage} is applied $L$-1 steps to obtain the embeddings $\{\textbf{h}_{1}^{L-1}, \cdots, \textbf{h}_{K}^{L-1}\}$ for the $K$ first-order neighbors of $u$. Then we aggregate them together to obtain the embedding of the target user $u$. Unlike the previous $L$-1 steps that concatenates  $\textbf{h}_{u}^{l-1}$ and  $\textbf{h}^{l}_{  \mathcal{N}(u)}$  to obtain $\textbf{h}_u^{l}$ for each neighbor (Cf. Eq.~\eqref{eq:GraphSage}), we only use $\textbf{h}^{L}_{  \mathcal{N}(u)}$ to represent the target embedding $\textbf{h}_u^{L}$, as we aim to predict the target embedding by the neighbors' embeddings:

\beqn{
	\label{eq:pre-training-aggregation}
	\textbf{h}^{L}_{\mathcal{N}(u)} &=& \text{AGGREGATE}(\{\textbf{h}^{L-1}_i, \forall i \in \mathcal{N}(u)\}),	\\ \nonumber
	\textbf{h}_u^{L} &=& \sigma ( \mathbf{W}^L \cdot  \textbf{h}_{\mathcal{N}(u)}^{L}  ).\\ \nonumber
}

Finally, we use cosine similarity to measure the difference between the predicted target embedding $\textbf{h}_u^{L}$ and the ground-truth embedding $\textbf{h}_u$, as proposed by~\cite{ziniuhufewshot19}, 
due to its popularity as an indicator for the semantic similarity between embeddings:

 \beqn{
	\label{eq:consine_similarity}
	\Theta_f^{*} &=& \mathop{\arg\max}_{\Theta_f} \sum_{u} {\rm cos}( \textbf{h}_u^{L}, \textbf{h}_u ),
}

\noindent where $\Theta_f = \{ \mathbf{W}^L, \Theta_{gnn} \} $ is the set of the parameters in $f$. 

Training GNNs in the meta-learning setting can explicitly reconstruct the user/item embeddings, making GNNs easily and rapidly being adapted to new cold-start users/items.
After the model is trained, for a new arriving cold-start user or item, based on the few first-order neighbors and the high-order neighbors, we can predict an accurate embedding for it. However, the basic pre-training GNN model doesn't specially address the cold-start neighbors. During the original graph convolution process, the inaccurate embeddings of the cold-start neighbors and the embeddings of other neighbors are equally treated and aggregated to represent the target user/item. Although some GNN models such as GrageSAGE or FastGCN filter neighbors before aggregating them, they usually follow the random or importance sampling strategies, which ignore the cold-start characteristics of the neighbors. Out of this consideration, we incorporate a meta aggregator and an adaptive neighbor sampler into the above basic pre-training GNN model. 

\subsection{Meta Aggregator}
\label{sec:meta_aggregator}

We propose the Meta Aggregator to deal with the cold-start neighbors. Suppose the target node is $u$ and one of its neighbor is $i$, if $i$ is interacted with sparse nodes, its embedding, which is inaccurate, will affect the embedding of $u$ when performing graph convolution by the GNN $f$. Although the cold-start issue of $i$ is dealt with when $i$ acts as another target node, embedding $i$, which is parallel to embedding $u$, results in a delayed effect on $u$' embedding. Thus, before training the GNN $f$, we train another function $g$ under the similar meta-learning setting as $f$. The meta learner $g$ learns an additional embedding for each node only based on its first-order neighbors, thus it can quickly adapt to new cold-start nodes and produce more accurate embeddings for them. The embedding produced by $g$ is combined with the original embedding at each convolution in $f$. 
Although both $f$ and $g$ are trained under the same meta-learning setting, $f$ is to tackle the cold-start target ndoes, but $g$ is to enhance the cold-start neighbors' embeddings.

Specifically, we instantiate $g$ as a self-attention encoder~\cite{vaswaniselfattention17}. For each user $u$, $g$ accepts the initial embeddings $\{\textbf{h}_{1}^{0}, \cdots, \textbf{h}_{K}^{0}\}$ of the $K$ first-order neighbors for $u$ as input, calculates the attention scores of all the neighbors to each neighbor $i$ of $u$, aggregates all the neighbors' embeddings according to the attention scores to produce the embedding $\textbf{h}_i$ for each $i$, and finally averages the  embeddings of all the neighbors to get the embedding $\tilde{\textbf{h}}_u$, named as the meta embedding of user $u$. The process is formulated as:

\beqn{
	\label{eq:meta_learner}
	\{ \textbf{h}_{1}, \cdots, \textbf{h}_{K} \}  &\leftarrow& \text{SELF\_ATTENTION} ( \{ \textbf{h}_{1}^{0}, \cdots, \textbf{h}_{K}^{0} \} ), \\\nonumber
	 \tilde{\textbf{h}}_u &=& \text{AVERAGE} (\{ \textbf{h}_{1}, \cdots, \textbf{h}_{K} \}).
}

The self-attention technique, which pushes the dissimilar neighbors further apart and pulls the similar neighbors closer together, can capture the major preference of the nodes from its neighbors.  
The same cosine similarity described in Eq.\eqref{eq:consine_similarity} is used as the loss function to measure the difference between the predicted meta embedding $\tilde{ \textbf{h}}_u$ and the ground truth embeding $\textbf{h}_u$. 
Once $g$ is learned, we add the meta embedding $\tilde{ \textbf{h}}_u$  into each graph convolution step of the GNN $f$ in Eq.~\eqref{eq:GraphSage}:

\beqn{
	\label{eq:meta_aggregator}
	\textbf{h}_u^{l} &=&\sigma (\mathbf{W}^{l} \cdot {\rm CONCAT} (\tilde{\textbf{h}}_{u},  \textbf{h}_{u}^{l-1}  ,  \textbf{h}^{l}_{ \mathcal{N}(u)} ), 
}

\noindent where the target embedding $\textbf{h}_{u}^{l-1}$ of the former step, the aggregated neighbor embedding  $\textbf{h}^{l}_{ \mathcal{N}(u)}$ of this step are learned following the basic pre-training GNN model. 
For a target user $u$, Eq.~\eqref{eq:meta_aggregator} is repeated $L$-1 steps to obtain the embeddings $\{\textbf{h}_{1}^{L-1}, \cdots, \textbf{h}_{K}^{L-1}\}$ for its $K$ first-order neighbors, Eq.~\eqref{eq:pre-training-aggregation} is also applied on them to get the final embedding $\textbf{h}_u^L$, and finally the same cosine similarity in Eq.~\eqref{eq:consine_similarity} is used to optimize the parameters of the meta aggregator, which includes the parameters $ \Theta_{f}$ of the basic pre-training GNN and  $\Theta_{g}$ of the meta-learner. The meta aggregator extends the original GNN graph convolution through emphasizing the representations of the cold-start neighbors in each convolution step, which can improve the final embeddings of the target users/items.

\subsection{The Adaptive Neighbor Sampler}
\label{sec:neighbor_sampler}

The proposed sampler does not make any assumption about what kind of neighbors are useful for the target users/items. Instead, it learns an adaptive sampling strategy according to the feedbacks from the pre-training GNN model.
To achieve this goal, we cast the task of neighbor sampler as a hierarchical Markov Decision Process (MDP)~\cite{zhanghrl19,Ryuichirl19}.
Specifically, we formulate the neighbor sampler as $L-1$ MDP subtasks where the $l$-th subtask indicates sampling the $l$-order neighors. The subtasks are performed sequentially by sampling from the second-order to $L$-order neighbors\footnote{Since the first-order neighbors are  crucial for depicting the user profile, we maintain all the first-order neighbors and only sample the high-order neighbors.}.
When the $l$-th subtask deletes all the neighbors or the $L$-th subtask is finished, the overall task is finished. 
We will introduce how to design the state, action and the reward for these subtasks as below.

\vpara{State.}  The $l$-th subtask  takes an action at the $t$-th $l$-order neighbor to determine whether to sample it or not according to the state of the target user $u$, the formerly selected neighbors, and the $t$-th $l$-order neighbor to be determined. We define the state features $\mathbf{s}^l_t$ for the $t$-th $l$-order neighbor as the cosine similarity and the element-wise product between its initial embedding and the target user $u$'s initial embedding, the initial embedding of each formerly selected neighbor by the $l$-1-th subtask and the average embedding of all the formerly selected neighbors respectively.

\vpara{Action and Policy.} We define the action $a^l_t \in \{ 0, 1 \}$ for the $t$-th $l$-order neighbor as a binary value to represent whether to sample the neighbor or not. We perform $a^l_t$ by the policy function $P$:

\beqn{
	\label{eq:policy_function}
	\textbf{H}_t^l &=&  \text{ReLU} (\textbf{W}_1^l\textbf{s}_t^l +\textbf{b}^l), \\ \nonumber
	P(a_t^l|\textbf{s}_t^l, \Theta^l_s) &=& a_t^l \sigma( \textbf{W}_2^l\textbf{H}_t^l) + (1-a_t^l) (1- \sigma (\textbf{W}_2^l\textbf{H}_t^l)),  \nonumber
}

\noindent where $\textbf{W}_1^l \in \mathbb{R}^{d_s \times d}$, $\textbf{W}^l_2 \in \mathbb{R}^{d \times 1}$ and $\textbf{b}^l \in \mathbb{R}^{d_s}$ are the parameters to be learned,  $d_s$ is the number of the state features and $d$ is the embedding size. Notation $\textbf{H}_t^l$ represents the embedding of the input state and $\Theta^l_s=\{\textbf{W}^l_1, \textbf{W}^l_2, \textbf{b}^l\}$. Sigmoid function $\sigma$ is used to transform the input state into a probability.

\vpara{Reward.}
The reward is a signal to indicate whether the performed actions are reasonable or not. Suppose the sampling task is finished at the $l'$-th subtask, each action of the formerly performed $l'$ subtasks accepts a delayed reward after the last action of the $l'$-level subtask. In another word, the immediate reward for an action is zero except the last action. The reward is formulated as:

\beq{ 
	\label{eq:reward}
	R(a_t^l, \textbf{s}_t^l) 
	\!=\! \left\{
	\begin{array}{cl}
		{\!\!\!
			\rm cos}( \hat{\textbf{h}}^L_u ,\!\textbf{h}_u)- {\rm cos}(\textbf{h}^L_u ,\!\textbf{h}_u) & \!\!\!\! \mbox{if } t = |\mathcal{N}^{l'}(u)| \wedge l = l';  \\
			0 & \!\!\!\!\mbox{otherwise,}
	   \end{array}\right. 
}
	
				\noindent where $\textbf{h}^L_u$ is the predicted embedding of the target user $u$ after the $L$-step convolution by Eq.~\eqref{eq:meta_aggregator} and Eq.~\eqref{eq:pre-training-aggregation}, while $\hat{\textbf{h}}^L_u$ is predicted in the same way but on the sampled neighbors following the policy function in Eq.~\eqref{eq:policy_function}. The cosine similarity between the predicted embedding and the ground truth embedding indicates the performance of the pre-training GNN Model. The difference between the performance caused by $\hat{\textbf{h}}^L_u$ and $\textbf{h}^L_u$ reflects the sampling effect.
				
		\vpara{Objective Function.}
		We find the optimal parameters of the policy function defined in Eq.~\eqref{eq:policy_function} by maximizing the expected reward $\sum_{\tau} P(\tau; \Theta_s) R(\tau)$, where $\tau=\{s_1^1, a_1^1, s_2^1,\cdots, s_t^{l'},a_t^{l'},s_{t+1}^{l'},\cdots\}$ is a sequence of the sampled actions and the transited states, $P(\tau; \Theta_s)$ denotes the corresponding sampling probability, $R(\tau)$ is the reward for the sampled sequence $\tau$, and $\Theta_s= \{   \Theta_s^1, \cdots,  \Theta^L_s \}$. 
		Since there are too many possible action-state trajectories for the entire sequence, we adopt the monto-carlo policy gradient~\cite{Williams92} to sample $M$ action-state trajectories and calculate the gradients:
		
		{\small \beq{
				\label{eq:gradient1}
				\nabla_{\Theta_s} = \frac{1}{M} \sum_{m=1}^M \sum_{l=1}^{l'}  \sum_{t=1}^{|\mathcal{N}^{l}(u)|}\nabla_{\Theta_s} \log P(a^{m,l}_{t}|\textbf{s}^{m,l}_{t}, \Theta^l_s)  R(a^{m,l}_{t},s^{m,l}_{t}) , 
		}}
		
		\noindent where $s^{m,l}_{t}$ represents the state of the $t$-th $l$-order neighbor in the $m$-th action-state trajectory, and $a^{m,l}_{t}$ denotes the corresponding action. $\mathcal{N}^{l}(u)$ indicates the set of all the $l$-order neighbors. 
		
		Algorithm~\ref{algo:rl2} shows the training process of the adaptive neighbor sampler. At each step $l$, we sample a sequence of actions $A^l$ (Line 5). If all the actions at the $l$-th step equal to zero or the last $L$-th step is performed (Line 6), the whole task is finished, then we compute the reward (Line 7) and the gradients (Line 8). After an epoch of sampling, we update the parameters of the sampler (Line 10). If it is jointly trained with the meta learner and the meta aggregator, we also update their parameters (Line 12).
   		
			\normalem
		\begin{algorithm}[t]
			{\small \caption{ The Joint Training Process.  \label{algo:rl2}}
				\KwIn{$Train_T = \{  (u_k,i_k)\}$, the ground truth embeddings $\{ (\textbf{h}_u, \textbf{h}_i) \}$, a pre-trained meta learner with $\Theta^0_g$, meta aggregator  with $\Theta^0_f$ and $\Theta^0_g$ and neighbor sampler  with  $\Theta^0_s$.}		
				Initialize $\Theta_s=\Theta^0_s$, $\Theta_f=\Theta^0_f$, $\Theta_g=\Theta^0_g$ \;
				\For{ epoch from 1 to  E }{
					\ForEach{$u_k$ or $i_k$ in $Train_T$}{
						\For{  $l$ in  $\{ 2, 3, \cdots, L\}$ }{ 
							Sample a sequence of actions $ \tau^l = \{a_1^l, \cdots, a_t^l, \cdots,a_{|\mathcal{N}^l(u)|}^l    \}$ by Eq.~\eqref{eq:policy_function}\;
							
							\If{$ \forall$  $a_t^l = 0$  or $l = L$}{
								Compute $R(a_{|\mathcal{N}^l(u)|}^l, \textbf{s}_{|\mathcal{N}^l(u)|}^l)$ by Eq.~\eqref{eq:reward}\;
								Compute gradients by Eq.~\eqref{eq:gradient1}\;
								Break;
							}
						}
					}
					Update $\Theta_s$\;
					\If{Jointly Training}{		
						Update $\Theta_g$ and $\Theta_f$ \;
					}
				}
			}
		\end{algorithm}
		\ULforem

		\subsection{Model Training} 
		The whole process of the pre-training GNN model is shown in Algorithm~\ref{algo:rl}, where we first pre-train the meta learner $g$ only based on first-order neighbors (Line 1),  and then incorporate $g$ into each graph convolution step to pre-train the meta aggregator (Line 2), next we pre-train the neighbor sampler with feedbacks from the pre-trained meta aggregator  (Line 3),  and finally we jointly train the meta learner, the meta aggregator and the neighbor sampler together (Line 4). 
		Same as the settings of \cite{zhanghrl19,junrl18}, to have a stable update during joint training, each parameter $\Theta \in \{ \Theta_{f}, \Theta_{g}, \Theta_{s}  \}$  is updated by a linear combination of its old version and the new old version, i.e., $\Theta_{new} = \lambda \Theta_{new} + (1-\lambda) \Theta_{old}$, where $\lambda \ll 1$.

			\begin{algorithm}[t]
			{\small \caption{The Overall Training Process. \label{algo:rl}}
				Pre-train the meta learner with parameter $\Theta_g$; \\
				Pre-train the meta aggregator  with parameter $\Theta_f$ when fixing $\Theta_g$;\\
				Pre-train the neighbor sampler with parameter $\Theta_s$ by Algorithm~\ref{algo:rl2} when fixing $\Theta_g$ and $\Theta_f$;\\
				Jointly train the three modules together with parameters $\Theta_g$, $\Theta_f$ and $\Theta_s$  by running Algorithm~\ref{algo:rl2}; \\
				
		}\end{algorithm}
		
		\subsection{Downstream Recommendation Task}
		\label{sec:fine-tune}
		After the pre-training GNN model is learned, we can fine-tune it in the recommendation downstream task. Specifically, for each target user $u$ and his  neighbors $ \{  \mathcal{N}^{1}(u), \cdots, \mathcal{N}^{L}(u)  \}$ of different order, we first use the pre-trained neighbor sampler to sample proper high-order neighbors  $\{  \mathcal{N}^{1}(u), \hat{\mathcal{N}}^{2}(u) \cdots, \hat{\mathcal{N}}^{L}(u)  \}$, and then use the pre-trained meta aggregator to produce the user embedding $\textbf{h}_u^{L}$. The item embeddings are generated in the same way. 
		Then we transform the embeddings and make a product between a user and an item to obtain the relevance score $y(u, i) = {\sigma ( \mathbf{W} \cdot  \textbf{h}_u^{L}  )}^\mathrm{T}  \sigma ( \mathbf{W} \cdot  \textbf{h}_i^{L}  )$ with parameters	$\Theta_r = \{ \mathbf{W}\}$. The  BPR loss defined in Eq.~\eqref{bprloss} is used to optimize $\Theta_r$ and fine-tune $\Theta_{g}$, $\Theta_{f}$ and $\Theta_{s}$.

\section{Experiment}

In this section, we present two types of experiments to evaluate the performance of the proposed pre-training GNN model. One is an intrinsic evaluation which aims to directly evaluate the quality of the user/item embedding predicted by the pre-training model. The other one is an extrinsic evaluation which applies the proposed pre-training model into the downstream recommendation task and indirectly evaluate the recommendation performance. 

\subsection{Experimental Setup}

\vpara{Dataset.}
We evaluate on three public datasets including MovieLens-1M (Ml-1M)\footnote{https://grouplens.org/datasets/movielens/}~\cite{harper2016movielens}, MOOCs\footnote{http://moocdata.cn/data/course-recommendation}~\cite{zhanghrl19} and Last.fm\footnote{http://www.last.fm}. Table~\ref{tb:statistics} illustrates the statistics of these datasets. The code is available now~\footnote{https://github.com/jerryhao66/Pretrain-Recsys}.

\begin{table}[t]
	\newcolumntype{?}{!{\vrule width 1pt}}
	\newcolumntype{C}{>{\centering\arraybackslash}p{4.6em}}
	\caption{
		\label{tb:statistics} Statistics of the Datasets.
		\normalsize
	}
	
	\centering  \small
	\renewcommand\arraystretch{1.0}
	
	\begin{tabular}{@{~}l@{~} @{~}r@{~} @{~}r@{~} @{~}r@{~} @{~}r@{~} }
		\toprule
		\multirow{2}{*}{\vspace{+0.36cm} Dataset}
		&\#Users&\#Items&\#Interactions & \#Sparse Ratio  
		\\
		\midrule
		MovieLens-1M
		&6,040 &3,706&1,000,209& 4.47\% 
		\\
		MOOCs
		& 82,535 & 1,302 & 458,453 	& 0.42\% 
		\\
		Last.fm
		&992&1,084,866&19,150,868& 1.78\%
		\\
		\bottomrule
	\end{tabular}
	
\end{table}

\vpara{Baselines.}
We select three types of baselines including the state-of-the-art neural matrix factorization model, the general GNN models and the special GNN models for recommendation:

\begin{itemize}[ leftmargin=10pt ]
\item \textbf{NCF~\cite{he2017neural}}: is a neural matrix factorization model which combines Multi-layer Perceptron and matrix factorization to learn the embeddings of users and items.

\item \textbf{GraphSAGE~\cite{williamgraphsage17}}: is a general GNN model which samples neighbors randomly and aggregates them by the AVERAGE function.

\item \textbf{GAT~\cite{gat18}}: is a general GNN model which aggregates neighbors by the attention mechanism without sampling.

\item \textbf{FastGCN~\cite{chenfastgcn18}}: is also a general GNN model which samples the neighbors by the important sampling strategy and aggregates neighbors by the same aggregator as GCN~\cite{Thomasgcn}.

\item \textbf{FBNE~\cite{chenhongxutked20}}: is a special GNN model for recommendation, which samples the neighbors by the importance sampling strategy and aggregates them by the AVERAGE function based on the explicit user-item and the implicit user-user/item-item interactions.

\item \textbf{LightGCN~\cite{xiangnanhe_lightgcn20}}: is  a special GNN model for recommendation, which discards the feature transformation and the nonlinear activation functions in the GCN aggregator.

\end{itemize}
 
 For each GNN model, we evaluate the corresponding pre-training model. For example, for the GAT model, Basic-GAT means we apply GAT into the basic pre-training GNN model proposed in Section~\ref{sec:basic_pretraining_gnn}, Meta-GAT indicates we incorporate the meta aggregator proposed in Section~\ref{sec:meta_aggregator} into Basic-GAT, NSampler-GAT represents that we incorporate the adaptive neighbor sampler proposed in Section~\ref{sec:neighbor_sampler} into Basic-GAT, and GAT* is the final poposed pre-training GNN model that incorporates both the meta aggregator and the adaptive neighbor sampler into Basic-GAT. 
 
 The original GAT and LightGCN models use the whole adjacency matrix, i.e., all the neighbors, in the aggregation function. To train them more efficiently, we implement them in the same sampling way as GraphSAGE, where we randomly sample at most 10 neighbors for each user/item. Then the proposed pre-training GNN model is performed under the sampled graph.

\vpara{Intrinsic and Extrinsic Settings.} 
We divide each dataset into the meta-training set $D_T$ and the meta-test set $D_N$. We train and evaluate the pre-training GNN model in the intrinsic user/item embedding inference task on $D_T$. Once the  model is trained, we fine-tune it in the extrinsic downstream recommendation task and evaluate it on $D_N$. 
We select the users/items from each dataset with sufficient interactions as the target users/items in $D_T$, as the intrinsic evaluation needs the true embeddings of users/items inferred from the sufficient interactions. 
Take the scenario of cold-start users as an example, we divide the users with the number of the direct interacted items more than $n_i$ into $D_T$ and leave the rest users into $D_N$.  We select $n_i$ as 60 and 20 for the dataset Ml-1M and MOOCs respectively. Since the users in Last.fm interact with too many items, we randomly sample 200 users, put 100 users into $D_T$ and leave the rest 100 users into $D_N$. For each user in $D_N$, we only keep its $K$-shot items to simulate the cold-start users. 
Similarly, for the cold-start item scenario, we divide the items with the number of the direct interacted users more than $n_u$ into $D_T$ and leave the rest items into  $D_N$, where $n_u$ is set as 60, 20 and 15 for MovieLens-1M, MOOCs and Last.fm respectively. In the intrinsic task, $K$ is set as 3 and 8, while in the extrinsic task, $K$ is set as 8. The embedding size $d$ is set as 256.  The number of the state features $d_s$ is set as 2819.

\subsection{Intrinsic Evaluations: Embedding Inference}
In this section, we conduct the intrinsic evaluation of inferring the embeddings of cold-start users/items by the proposed pre-training GNN model. Both the evaluations on the user embedding inference and the item embedding inference are performed.

\vpara{Training and Test Settings.} We use the meta-training set $D_T$ to perform the intrinsic evaluation.
Specifically, we randomly split $D_T$ into the training set $Train_T$ and the test set $Test_T$ with a ratio of 7:3. 
We train NCF~\cite{he2017neural} to get the ground-truth embeddings for the target users/items in both $Train_T$ and $Test_T$\footnote{ We concatenate the embeddings produced by both the MLP and the GMF modules in NCF as the ground-truth embedding.}.  
To mimic the cold-start users/items on $Test_T$, we randomly keep $K$ neighbors for each user/item, which results in at most $K^l$ neighbors ($1 \leq l \leq 3$) for each target user/item. Thus $Test_T$ is changed into $Test_T'$.

The original GNN models are trained by BPR loss in Eq.~\eqref{bprloss} on $Train_T$. 
The proposed pre-training GNN models are trained by the cosine similarity in Eq.~\eqref{eq:consine_similarity} on $Train_T$. 
The NCF model is trained transductively to obtain the user/item embeddings on the merge dataset of $Train_T$ and $Test_T'$. 
The embeddings in both the proposed models and the GNN models are initialized by the NCF embedding results. 
We use Spearman correlation~\cite{ziniuhufewshot19} to measure the agreement between the ground truth embedding and the predicted embedding.

\hide{
In our implementation, $K$ is set as 3 and 8. 
The embedding size $d$ is set as 256. 
The number of the state features $d_s$ is set as 2819.
The learning rate for the basic pre-training GNN model, the meta aggregator, the adaptive neighbor sampler and the jointly training  is set as 0.005, 0.003, 0.003 and 0.001 respectively.
}
 


\begin{table*}[t]
	\newcolumntype{?}{!{\vrule width 1pt}}
	\newcolumntype{C}{>{\centering\arraybackslash}p{2.6em}}
	\caption{
		\label{tb:ncf_pretrain} Overall performance of user/item embedding inference (Spearman correlation). The layer depth $L$ is 3.
		\normalsize
	}
	\centering  \small
	\renewcommand\arraystretch{1.0}
	\begin{tabular}{@{~}l@{~}?*{1}{CC?}*{1}{CC?}*{1}{CC?}*{1}{CC?}  *{1}{CC?}*{1}{CC} }
		\toprule
		\multirow{2}{*}{\vspace{-0.3cm} Methods }
		&\multicolumn{2}{c?}{Ml-1M (user)}
		&\multicolumn{2}{c?}{MOOCs (user)}
		&\multicolumn{2}{c?}{Last.fm (user)}
		&\multicolumn{2}{c?}{Ml-1M (item)}
		&\multicolumn{2}{c?}{MOOCs (item)}
		&\multicolumn{2}{c}{Last.fm (item)}
		\\
		\cmidrule{2-3} \cmidrule{4-5} \cmidrule{6-7} \cmidrule{8-9} \cmidrule{10-11} \cmidrule{12-13} 
		& {3-shot} & {8-shot} & {3-shot} & {8-shot} &{3-shot} &{8-shot}   &{3-shot}&{8-shot}  &{3-shot}&{8-shot} &{3-shot}&{8-shot} \\
		\midrule 
		NCF
		& -0.017 &0.063
		&-0.098 &-0.062
		&0.042 &0.117
		&-0.118& -0.017
		&-0.036 &0.027
		&-0.036 &-0.018
		\\	
		\midrule
		GraphSAGE
		&0.035&0.105
		&0.085&0.128
		&0.104&0.134
		&0.113&0.156
		&0.116&0.182
		&0.112&0.198
		\\
		Basic-GraphSAGE
		& 0.076& 0.198
		& 0.103& 0.152
		&0.132 &0.184
		&0.145 &0.172
		&0.172 &0.196
		&0.166 & 0.208
		\\
		Meta-GraphSAGE
		&0.258&0.271
		&0.298&0.320
		&0.186&0.209
		&0.434&0.448
		&0.288&0.258
		&0.312&0.333
		\\
		NSampler-GraphSAGE
		&0.266& 0.284
		&0.294&0.336
		&0.196&0.212
		&0.448&0.460
		&0.286&0.306
		&0.326&0.336
		\\
		GraphSAGE*
		&\textbf{0.368}&\textbf{0.375}
		&\textbf{0.302}&\textbf{0.338}
		&\textbf{0.326}&\textbf{0.384}
		&\textbf{0.470}&\textbf{0.491}
		&\textbf{0.316}&\textbf{0.336}
		&\textbf{0.336}&\textbf{0.353}
		\\
		\midrule
		GAT
		&0.020&0.049
		&0.092&0.138
		&0.092&0.125
		&0.116&0.126
		&0.108&0.118
		&0.106&0.114
		\\
		Basic-GAT
		&0.046& 0.158
		&0.104 & 0.168
		&0.158 &0.180
		&0.134 &0.168
		&0.112 &0.126
		&0.209 & 0.243
		\\
		Meta-GAT
		&0.224&0.282
		&0.284&0.288
		&0.206	& 0.212
		&0.438&0.462
		&0.294&0.308
		&0.314&0.340
		\\
		NSampler-GAT
		&0.296&0.314
		&0.339&0.354
		&0.198&0.206
		&0.464&0.472
		&0.394&0.396
		&0.338&0.358
		\\
		GAT*
		&\textbf{0.365}&\textbf{0.379}
		&\textbf{0.306}&\textbf{0.366}
		&\textbf{0.309}& \textbf{0.394}
		&\textbf{0.496}&\textbf{0.536}
		&\textbf{0.362}&\textbf{0.384}
		&\textbf{0.346}&\textbf{0.364}
		\\
		\midrule
		FastGCN
		&0.009&0.012
		&0.063&0.095
		&0.082&0.114
		&0.002&0.036
		&0.007&0.018
		&0.007&0.013
		\\
		Basic-FastGCN
		&0.082&0.146
		&0.083&0.146
		&0.104&0.149
		&0.088& 0.113
		&0.099&0.121
		&0.159&0.182
		\\
		Meta-FastGCN
		&0.181&0.192
		&0.282&0.280
		&0.224&0.274
		&0.216&0.266
		&0.248&0.278
		&0.230&0.258
		\\
		NSampler-FastGCN
		&0.188&0.194
		&0.281&0.286
		&0.226&0.277
		&0.268&0.288
		&0.267&0.296
		&0.246&0.253
		\\
		FastGCN*
		&\textbf{0.198}&\textbf{0.212}
		&\textbf{0.288}&\textbf{0.291}
		&\textbf{0.266}&\textbf{0.282}
		&\textbf{0.282}&\textbf{0.298}
		&\textbf{0.296}&\textbf{0.302}
		&\textbf{0.268}&\textbf{0.278}
		\\
		\midrule
		FBNE
		&0.034&0.102
		&0.053&0.065
		&0.142&0.164
		&0.168&0.190
		&0.137&0.168
		&0.127&0.133
		\\
		Basic-FBNE
		&0.162&0.190
		&0.162&0.185
		&0.135&0.180
		&0.176&0.209
		&0.157&0.180
		&0.167&0.173
		\\
		Meta-FBNE
		&0.186&0.204
		&0.269&0.284
		&0.175&0.192
		&0.426&0.449
		&0.236&0.272
		&0.178&0.182
		\\
		NSampler-FBNE
		&0.208&0.216
		&0.259&0.283
		&0.203&0.207
		&0.422&0.439
		&0.226&0.273
		&0.164&0.183
		\\
		FBNE*
		&\textbf{0.242}&\textbf{0.265}
		&\textbf{0.306}&\textbf{0.321}
		&\textbf{0.206}&\textbf{0.219}
		&\textbf{0.481}&\textbf{0.490}
		&\textbf{0.301}&\textbf{0.382}
		&\textbf{0.182}&\textbf{0.199}
		\\
		\midrule
		LightGCN
		&0.093&0.108
		&0.060&0.068
		&0.162&0.184
		&0.201&0.262
		&0.181&0.232
		&0.213&0.245
		\\
		Basic-LightGCN
		&0.178&0.192
		&0.212&0.226
		&0.182&0.192
		&0.318&0.336
		&0.234&0.260
		&0.252&0.290
		\\
		Meta-LightGCN
		&0.226&0.241
		&0.272&0.285
		&0.206&0.221
		&0.336&0.346
		&0.314&0.331
		&0.372&0.392
		\\
		NSampler-LightGCN
		&0.238&0.256
		&0.286&0.294
		&0.204&0.212
		&0.348&0.384
		&0.296&0.314
		&0.356&0.401
		\\
		LightGCN*
		&\textbf{0.270}&\textbf{0.286}
		&\textbf{0.292}&\textbf{0.309}
		&\textbf{0.229}&\textbf{0.234}
		&\textbf{0.382}&\textbf{0.408}
		&\textbf{0.334}&\textbf{0.353}
		&\textbf{0.386}&\textbf{0.403}
		\\
		\bottomrule
	\end{tabular}
	
\end{table*}

\begin{figure*}[t]
	\centering
	\mbox{ 
		\subfigure[ MovieLens (user)]{\label{subfig:movielens-user}
			\includegraphics[width=0.13\textwidth]{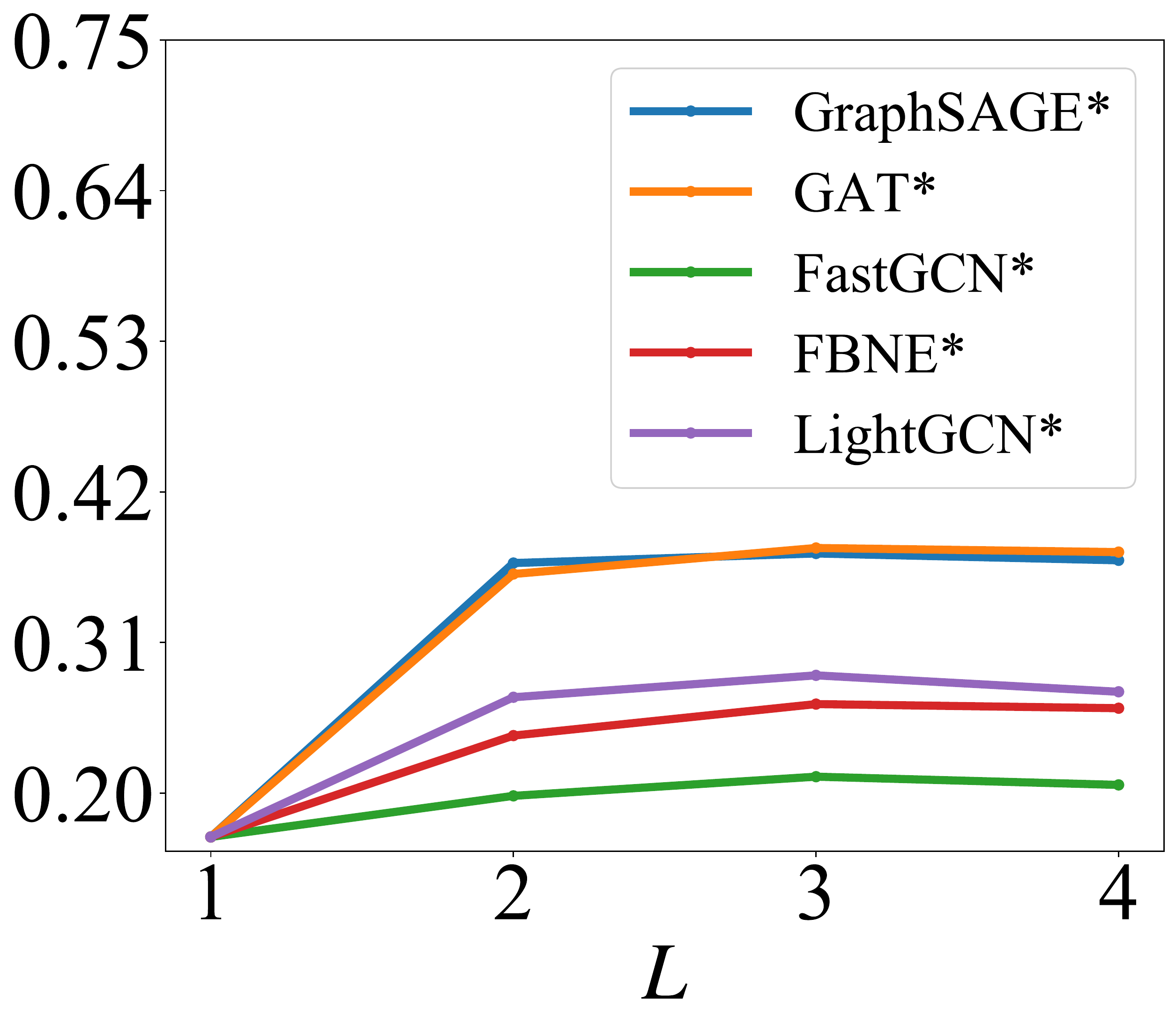}
		}

		\subfigure[ MovieLens (item)]{\label{subfig:movielens-item}
			\includegraphics[width=0.13\textwidth]{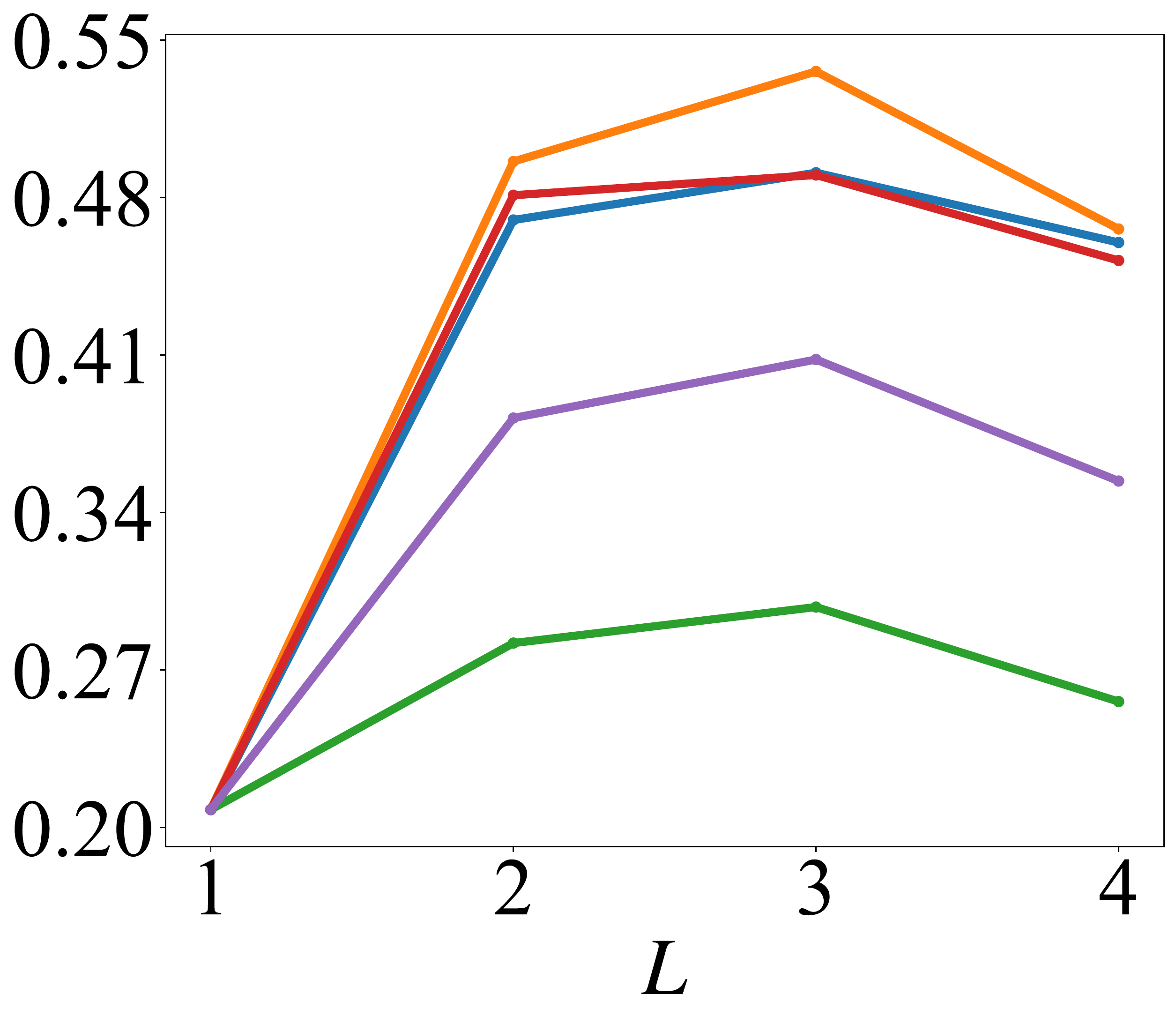}
		}
		\subfigure[ Moocs (user)]{\label{subfig:mooc-user}
			\includegraphics[width=0.13\textwidth]{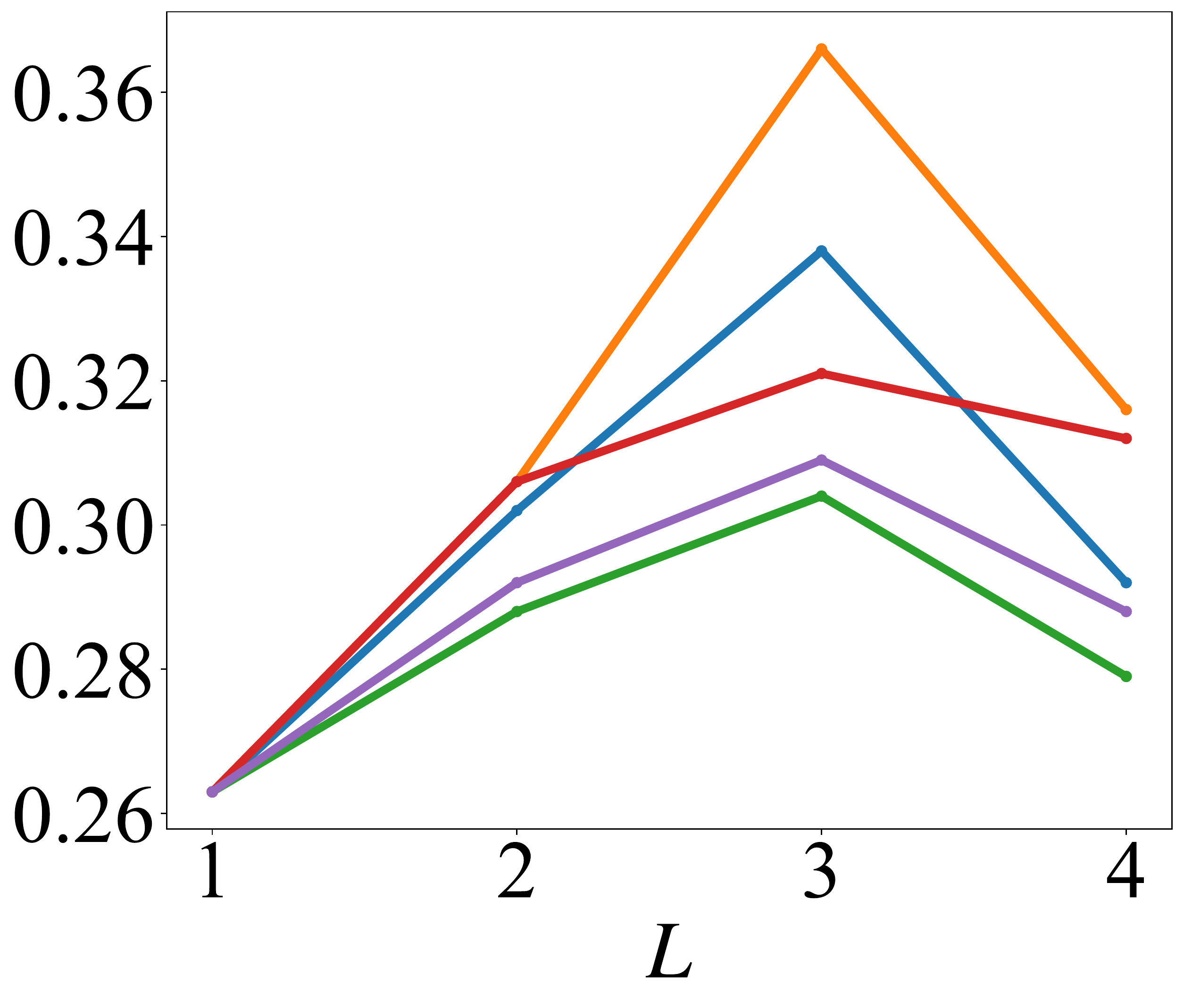}
		}

		\subfigure[ Moocs (item)]{\label{subfig:mooc-item}
			\includegraphics[width=0.13\textwidth]{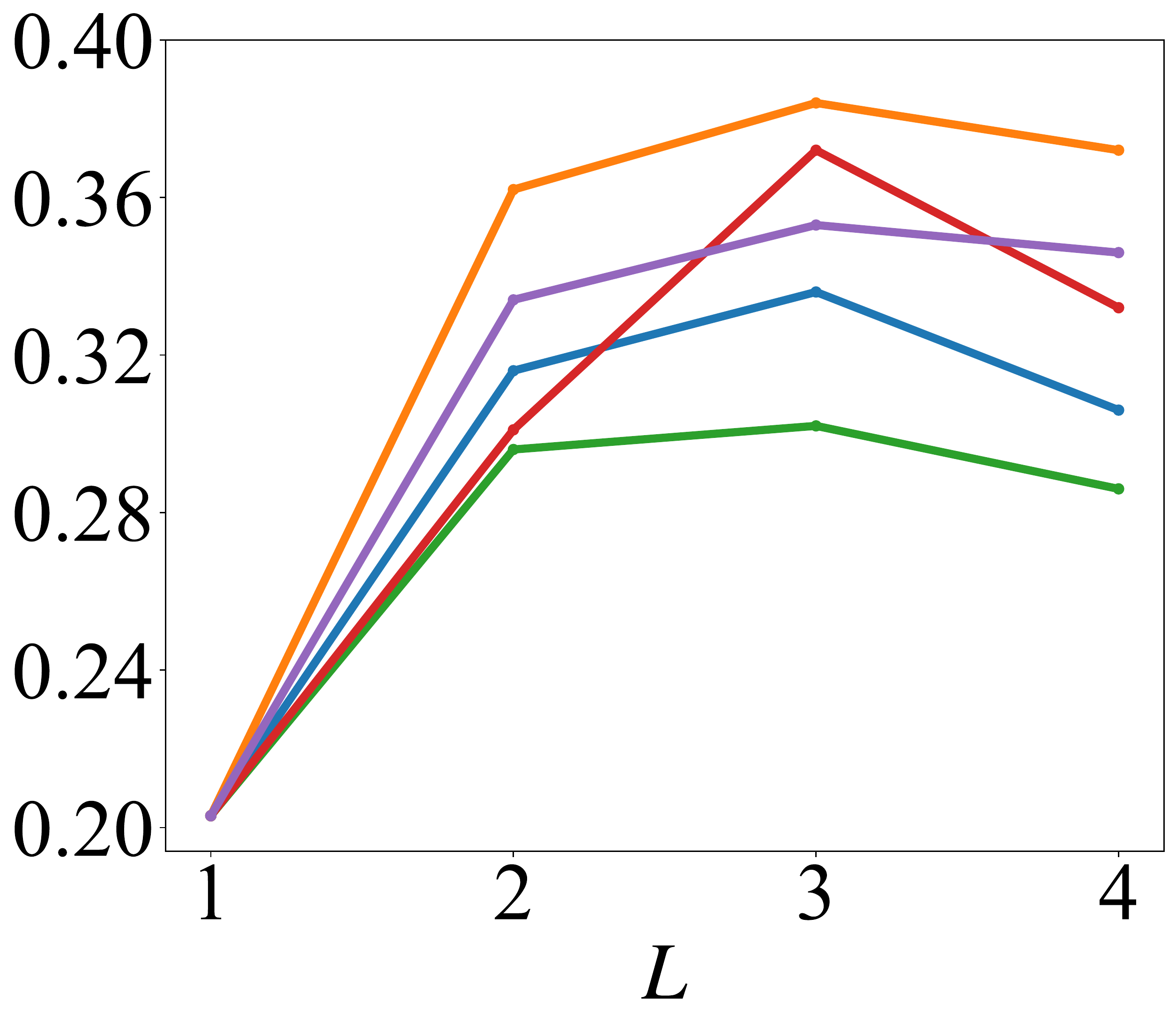}
		}
		\subfigure[ Last.fm (user)]{\label{subfig:last-user}
			\includegraphics[width=0.13\textwidth]{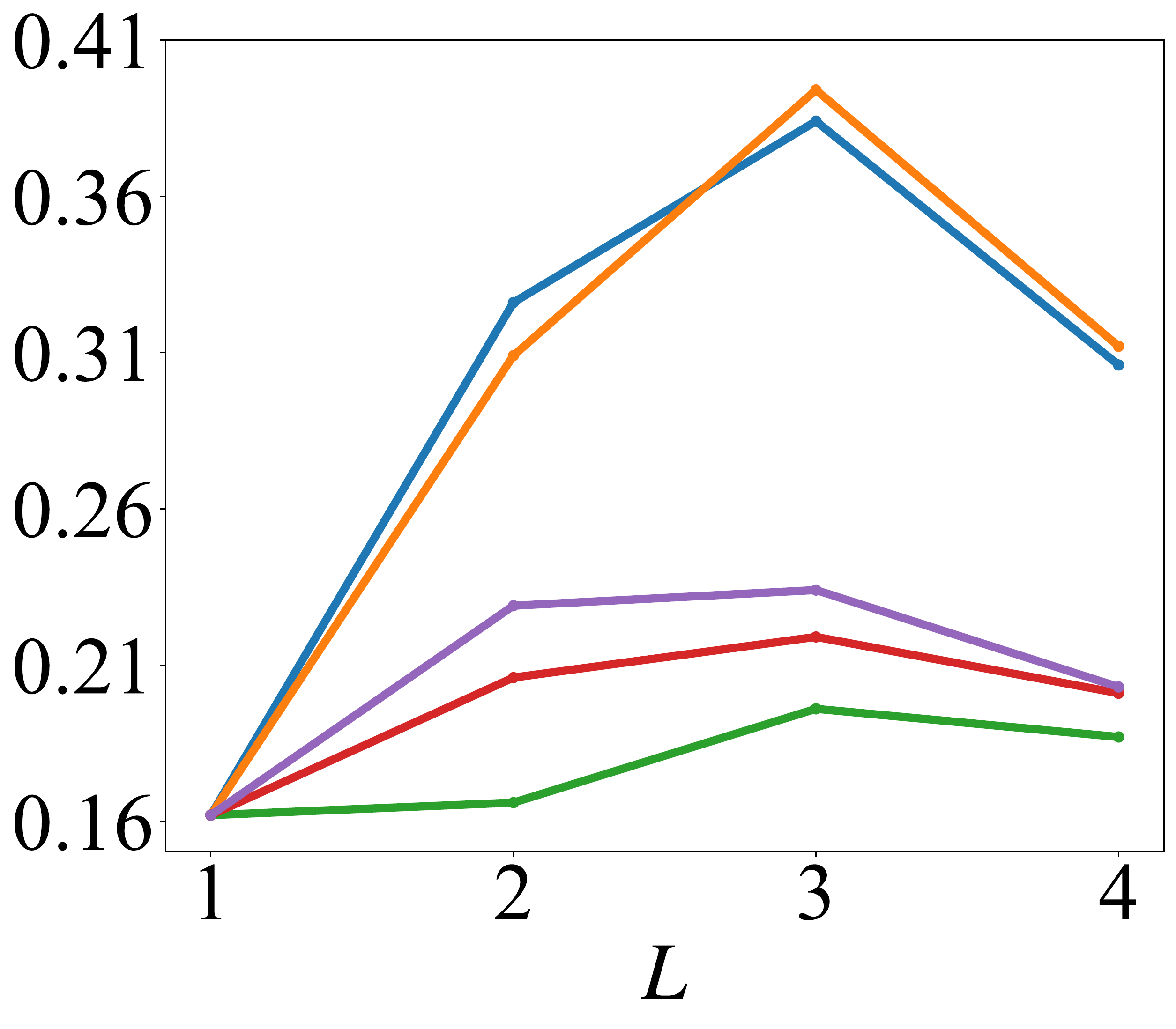}
		}

		\subfigure[ Last.fm (item)]{\label{subfig:lastfm-item}
			\includegraphics[width=0.13\textwidth]{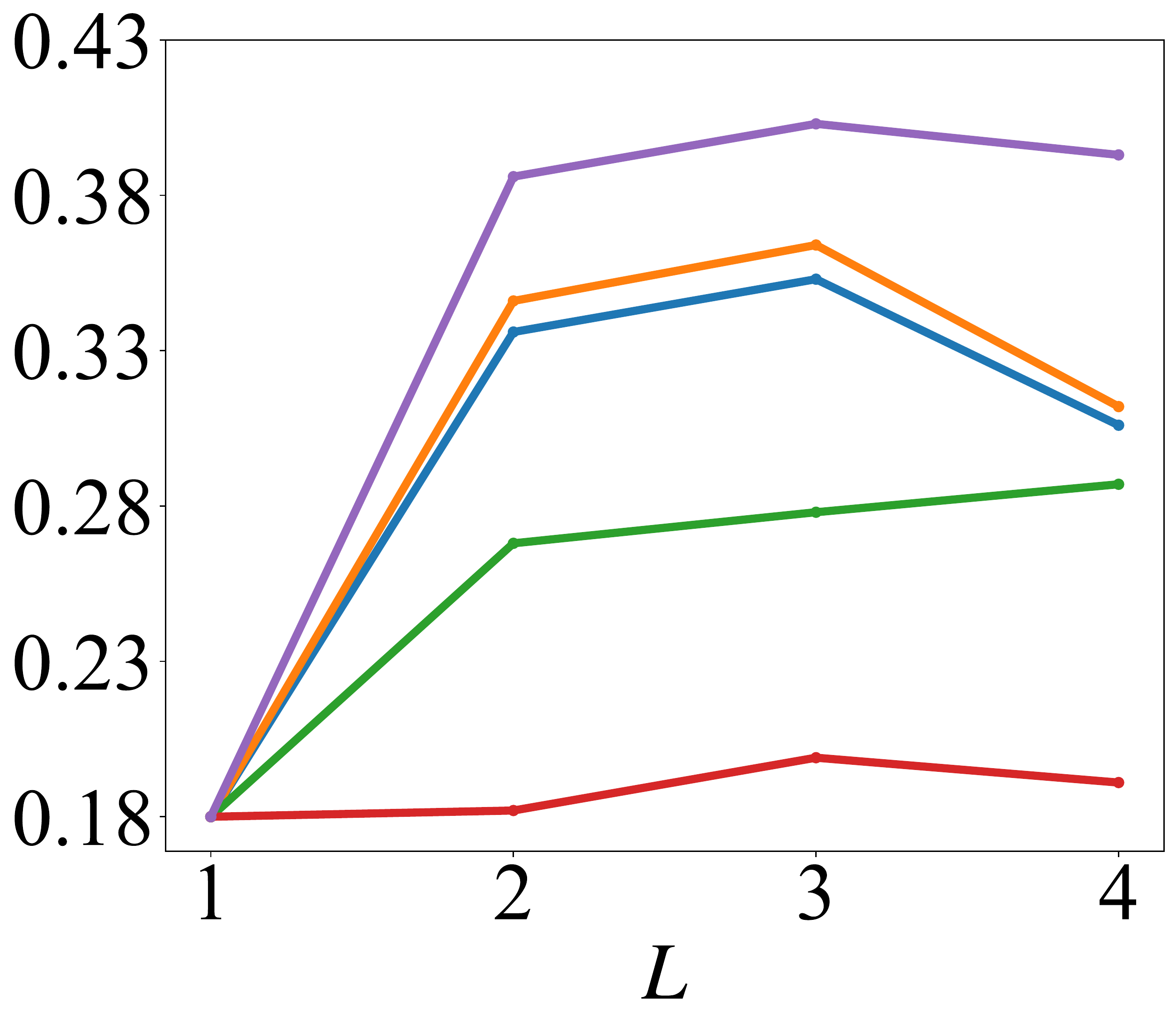}
		}

	}
	
	\caption{\label{fig:layer} Performance of user/item embedding inference under different layer depth $L$ when $K$=3. }
\end{figure*}

\vpara{Overall Performance.} Table~\ref{tb:ncf_pretrain} shows the overall performance of the proposed pre-training GNN model and all the baselines using 3-order neighbors. The results show that compared with the baselines, our proposed pre-training GNN model significantly improves the quality of the user/item embeddings (+33.9\%-58.4\% in terms of Spearman correlation). Besides, we have the following findings: 

\begin{itemize}[ leftmargin=10pt ]
\item Through incorporating the high-order neighbors,  the GNN models can improve the embedding quality of the cold-start users/items compared with the NCF model (+2.6\%-24.9\% in terms of Spearman correlation). 

\item  All the basic pre-training GNN models beat the corresponding GNN models by improving 1.79-15.20\% Spearman correlation, which indicates the basic pre-training GNN model is capable of reconstructing the cold-start user/item embeddings.

\item Compared with the basic pre-training GNN model, both the meta aggregator and the adaptive neighbor sampler can improve the embedding quality by 1.09\%-18.20\% in terms of the Spearman correlation, which indicates that the meta aggregator can indeed strengthen each layer's aggregation ability and the neighbor sampler can filter out the noisy neighbors.
 
\item When the neighbor size $K$ decreases from 8 to 3, the Spearman correlation of all the baselines significantly decrease 0.08\%-6.10\%, while the proposed models still keep a competitive performance. 

\item We also investigate the effect of the propagation layer depth $L$ on the model performance. In particular, we set the layer depth $L$ as 1,2,3 and 4, and report the performance in Fig.~\ref{fig:layer}. The results show when $L$ is 3, most algorithms can achieve the best performance, while only using the first-order neighbors performs the worst\footnote{When $L$ is 1, the neighbor sampler and the meta aggregator are not used, thus all the models get the same performance.}, which implies incorporating proper number of layers can alleviate the cold-start issue. 

\hide{
\item Among all the GCN methods, we find FastGCN performs the worst, even worse than the random sampling strategy GraphSAGE. The reason is that the representation of the target cold-start user/item is not accurate, which may disturb sampling proper high-order neighbors \textcolor{red}{But GCN also has the limitation}. The performance of the GCN-based recommendation method FBNE and LightGCN is better than other baselines, as FBNE explicitly consider user-user and item-item relationship, and LightGCN remove the transformation and nonlinear operation during aggregation operation. However, the cold-start problem is not thoroughly solved by these methods, because cold-start neighbors aren't dealt with specially when aggregating the embeddings of the neighbors for the cold-start users/items, while our proposed pre-training framework can solve this problem.  \textcolor{red}{What do you want to emphasize?}  }

\end{itemize}

\subsection{Extrinsic Evaluation: Recommendation}

In this section, we apply the pre-training GNN model into the downstream recommendation task and evaluate the  performance. 

\vpara{Training and Testing Settings.}
We consider the scenario of the cold-start users and use the meta-test set $D_N$ to perform recommendation. For each user in $D_N$, we select top 10\% of his interacted items in chronological order into the training set $Train_N$, and leave the rest items into the test set $Test_N$. 
We pre-train our model on $D_T$ and fine-tune it on $Train_N$ according to \secref{sec:fine-tune}. 

The original GNN and the NCF models are trained by the BPR loss function in Eq.~\eqref{bprloss} on $D_T$ and $Train_N$. For each user in $Test_N$, we calculate the user's relevance score to each of the rest 90\% items. 
We adopt Recall@$\mathcal{K}$ and NDCG@$\mathcal{K}$ as the metrics to evaluate the items ranked by the relevance scores. By default, we set $\mathcal{K}$ as 20 for Ml-1m and Moocs. For Last.fm, since there are too many items, we set $\mathcal{K}$ as 200.  
\hide{In our implementation, the neighbor size $K$ is set as 8,  the embedding size $d$ is set as 256 and the learning rate for the recommendation model is set as 0.003. }


\vpara{Overall Performance.} Table ~\ref{tb:recommendation} shows the overall recommendation performance. The results indicate that the proposed basic pre-training GNN models outperform the corresponding original GNN models by 0.40\%-3.50\% in terms of NDCG, which demonstrates the effectiveness of the basic pre-training GNN model on the cold-start recommendation performance. Upon the basic pre-training model, adding the meta aggregator and the adaptive neighbor sampler can further improve 0.30\%-6.50\% NDCG respectively, which indicates the two components can indeed alleviate the impact caused by the cold-start neighbors when embedding the target users/items, thus they can improve the downstream recommendation performance.  

\begin{figure}[t]
	\centering
	\includegraphics[width= 0.43 \textwidth]{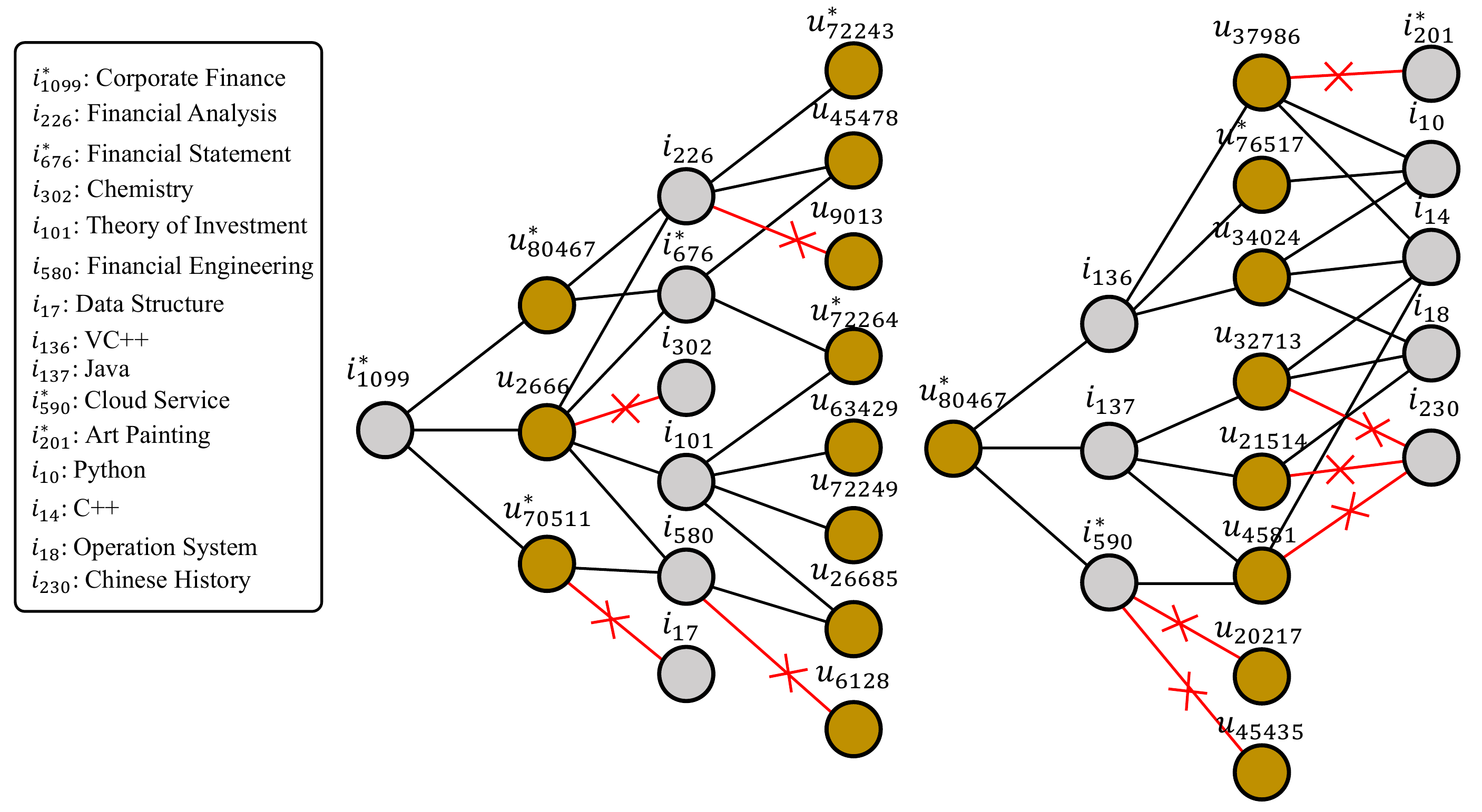}
	\caption{\label{fig:case_study} Case study of the adaptive neighbor sampling. }
\end{figure}

\vpara{Case Study.} We attempt to understand how the proposed pre-training model samples the high-order neighbors of the cold-start users/items by the MOOCs dataset.  Fig.~\ref{fig:case_study} illustrates two sampling cases, where notation * indicates the users/items are cold-start. 
The cold-start item   $i^*_{1099}$ is ``Corporate Finance", which only interacts with three users.  Our proposed neighbor sampler samples a second-order item $i^*_{676}$, ``Financial Statement". Although $i^*_{676}$ only interacts with two users, it is relevant to the target item $i^*_{1099}$. Similarly, the cold-start user $u^*_{80467}$, who likes computer science, only selects three computer science related courses. The proposed neighbor sampler samples a second-order user $u^*_{76517}$. Although it only interacts with two courses, they are ``Python" and ``VC++" which are relevant to computer science. 
However, the importance-based sampling strategies in FastGCN and FBNE cannot sample these neighbors, as they are cold-start with few interactions.

\begin{table}[t]
	\newcolumntype{?}{!{\vrule width 1pt}}
	\newcolumntype{C}{>{\centering\arraybackslash}p{2.4em}}
	\caption{
		\label{tb:recommendation} The Overall Recommendation Performance.
		\normalsize
	}
	\centering  \small
	\renewcommand\arraystretch{1.0}
	\begin{tabular}{@{~}l@{~}?*{1}{CC?}*{1}{CC?}*{1}{CC} @{}}
		\toprule
		\multirow{2}{*}{\vspace{-0.3cm} Methods }
		&\multicolumn{2}{c?}{Ml-1M}
		&\multicolumn{2}{c?}{MOOCs}
		&\multicolumn{2}{c}{Last.fm}
		\\
		\cmidrule{2-3} \cmidrule{4-5} \cmidrule{6-7}
		& {Recall} & {NDCG} & {Recall} & {NDCG} &{Recall} &{NDCG} \\
		\midrule 
		NCF
		&0.008&0.101 
		&0.021 & 0.047 
		&0.005 & 0.007
		\\
		\midrule
		GraphSAGE
		&0.006 &0.082
		&0.085 &0.066
		&0.003 &0.011
		\\
		Basic-GraphSAGE
		&0.013&0.135
		&0.082&0.091
		&0.007&0.044
		\\
		Meta-GraphSAGE
		&0.016&0.209 
		&0.096&0.116 
		&0.007&0.097
		\\
		NSampler-GraphSAGE
		&0.021&0.221 
		&0.101&0.122 
		&0.008 &0.088
		\\
		GraphSAGE*
		&\textbf{0.024}&\textbf{0.235}
		&\textbf{0.110}&\textbf{0.129}
		&\textbf{0.008}&\textbf{0.131}
		\\
		\midrule
		GAT
		&0.008&0.099 
		&0.023&0.055
		&0.006&0.033
		\\
		Basic-GAT
		&0.016&0.163
		&0.032&0.093
		&0.006&0.147
		\\
		Meta-GAT
		&0.017&0.191
		&0.063&0.123
		&0.009&0.184
		\\
		NSampler-GAT
		&0.012&0.188
		&0.084&0.132
		&0.010&0.199
		\\
		GAT*
		&\textbf{0.014}&\textbf{0.208}
		&\textbf{0.100}&\textbf{0.139}
		&\textbf{0.018}&\textbf{0.232}
		\\
		\midrule
		FastGCN
		&0.003&0.019
		&0.064&0.089
		&0.006&0.068
		\\
		Basic-FastGCN
		&0.008&0.102
		&0.099&0.117
		&0.012&0.083
		\\
		Meta-FastGCN
		&0.009&0.123
		&0.105&0.124
		&0.018&0.116
		\\
		NSampler-FastGCN
		&0.011&0.118
		&0.108&0.128
		&0.020&0.136
		\\
		FastGCN*
		&\textbf{0.012}&\textbf{0.123}
		&\textbf{0.119}&\textbf{0.140}
		&\textbf{0.023}&\textbf{0.186}
		\\
		\midrule
		FBNE
		&0.002&0.088
		&0.048&0.041
		&0.009&0.013
		\\
		Basic-FBNE
		&0.009&0.104
		&0.064&0.087
		&0.003&0.032
		\\
		Meta-FBNE
		&0.012&0.101
		&0.088&0.101
		&0.006&0.087
		\\
		NSampler-FBNE
		&0.013&0.118
		&0.102&0.117
		&0.006&0.099
		\\
		FBNE*
		&\textbf{0.014}&\textbf{0.121}
		&\textbf{0.117}&\textbf{0.138}
		&\textbf{0.007}&\textbf{0.129}
		\\
		\midrule
		LightGCN
		&0.014&0.207
		&0.102&0.112
		&0.001&0.083
		\\
		Basic-LightGCN
		&0.012&0.211
		&0.112&0.121
		&0.005&0.097
		\\
		Meta-LightGCN
		&0.018&0.221
		&0.120&0.139
		&0.005&0.101
		\\
		NSampler-LightGCN
		&0.020&0.218
		&0.116&0.132
		&0.007&0.106
		\\
		LightGCN*
		&\textbf{0.022}&\textbf{0.227}
		&\textbf{0.123}&\textbf{0.142}
		&\textbf{0.007}&\textbf{0.114}
		\\
		\bottomrule
	\end{tabular}
	
\end{table}

\hide{
\begin{figure}[t]
	\centering
	\mbox{ 

		\subfigure[ MovieLens]{\label{subfig:movielens-item-recommender}
			\includegraphics[width=0.14\textwidth]{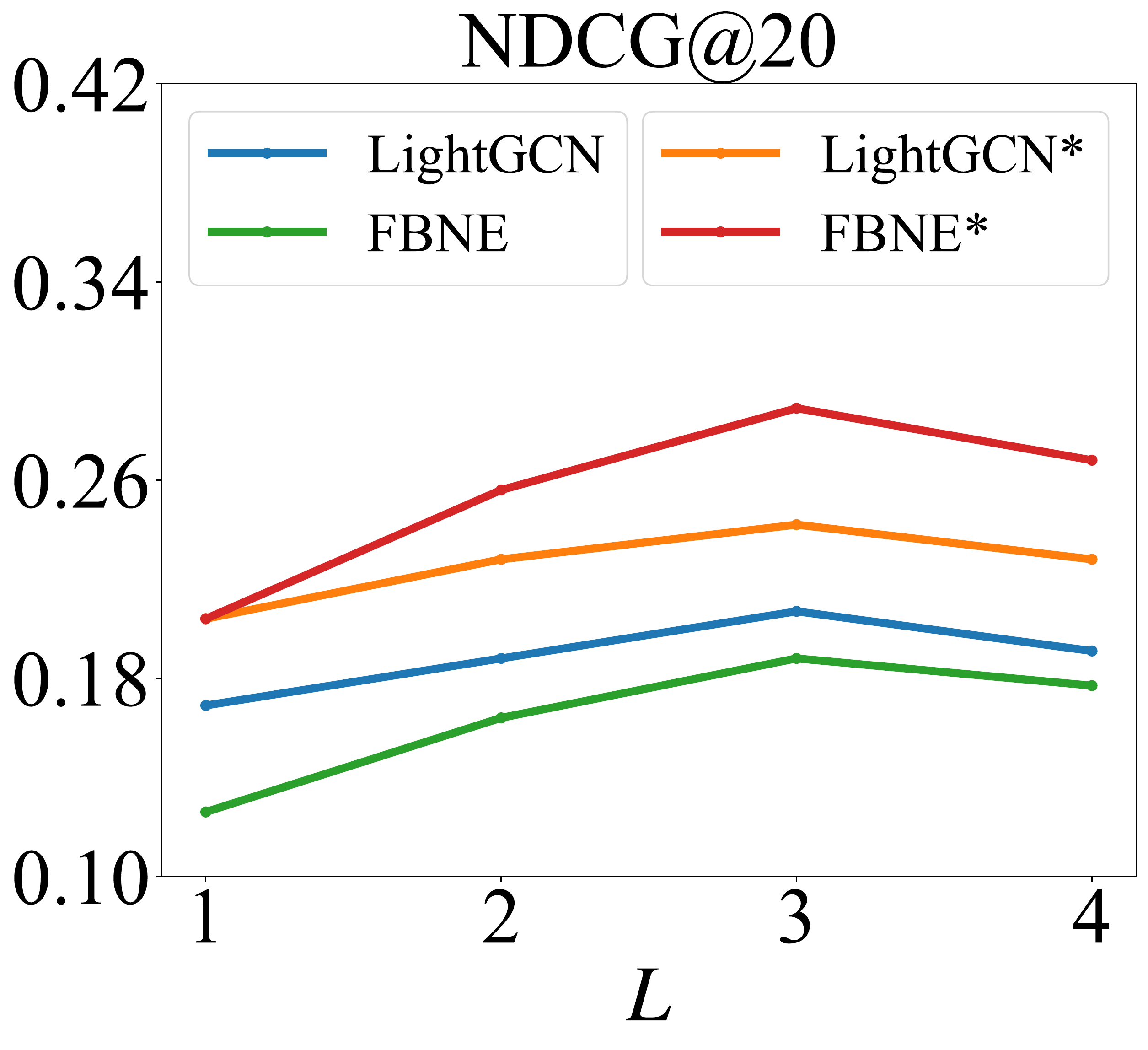}
		}

		\subfigure[ Moocs]{\label{subfig:mooc-item-recommender}
			\includegraphics[width=0.14\textwidth]{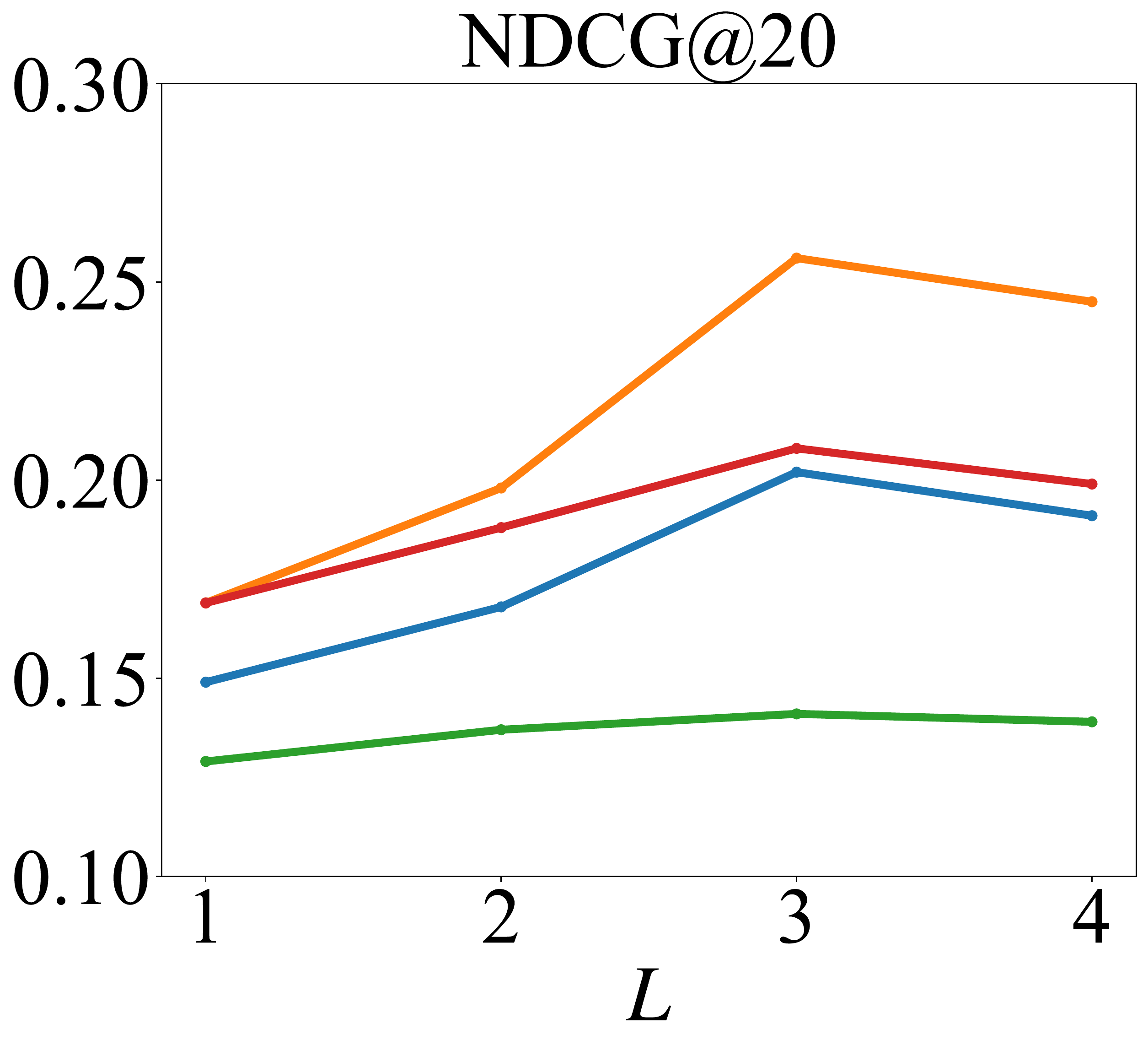}
		}

		\subfigure[ Last.fm]{\label{subfig:lastfm-item-recommender}
			\includegraphics[width=0.14\textwidth]{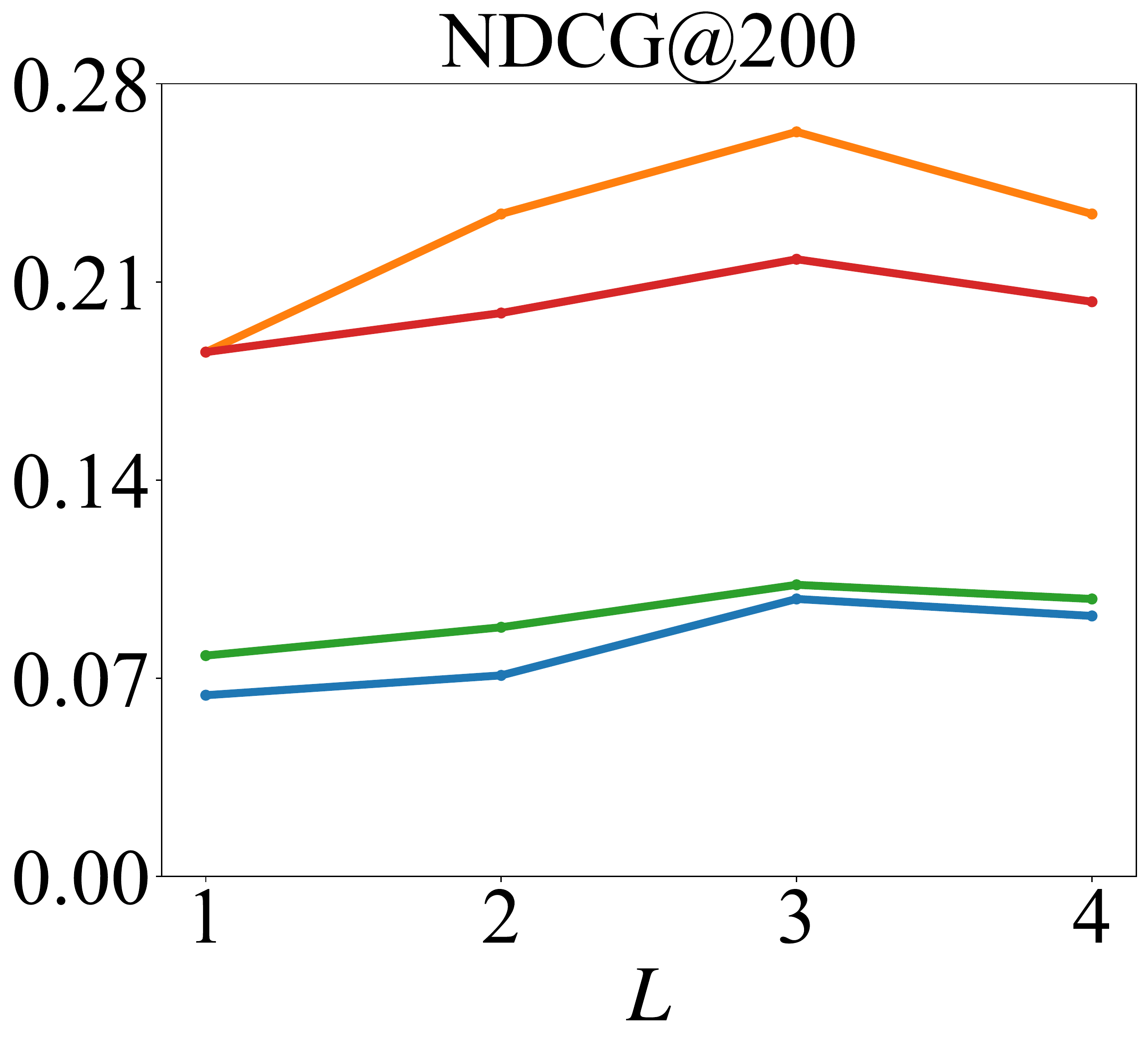}
		}
		
	}
	
	\caption{\label{fig:downstream_figure} Recommendation Performance under different layer depth $L$. }
\end{figure}
}

\section{Related Work}


\vpara{Cold-start Recommendation.}
Cold-start issue is a fundamental challenge in recommender systems.
On one hand, existing recommender systems incorporate the side information such as spatial information~\cite{YinWWCZ17,hongzhi_side}, social trust path~\cite{YinW0LYZ19,hongzhi_tkde_20,WangYWNHC19,GharibshahZHC20} and knowledge graphs~\cite{wang2019multi,wang2019kgat} to enhance the representations of the cold-start users/items. However, the side information is not always available, making it intractable to improve the cold-start embedding's quality.
On the other hand, researchers solve the cold-start issue by only mining the underlying patterns behind the user-item interactions. One kind of the methods is meta-learning~\cite{Finnmaml17,munkhdalaimetaneworks17,vinyalsmatching16,Snellprotocal17}, which consists of metric-based recommendation~\cite{vartak2017meta} and model-based recommendation~\cite{du2019sequential,lee2019melu,lu2020meta,pan2019warm}. However, few of them capture the high-order interactions. Another kind of method is GNNs, which leverage user-item bipartite graph to capture high-order collaborative signals for recommendation. The representative models include Pinsage~\cite{pinsage}, NGCF~\cite{wangncgf19}, LightGCN~\cite{xiangnanhe_lightgcn20}, FBNE~\cite{chenhongxutked20} and CAGR~\cite{hongzhi_tkde_20}. Generally, the recommendation-oriented GNNs optimize the likelihood of a user adopting an item,  which isn't a direct improvement of the embedding quality of the cold-start users or items.

\vpara{Pre-training GNNs.} 
Recent advances on pre-training GNNs aim to empower GNNs to capture the structural and semantic properties of an input graph, so that it can easily generalize to any downstream tasks with a few fine-tuning steps on the graphs~\cite{gpt_gnnhu}. 
The basic idea is to design a domain specific pretext task 
to provide additional supervision for exploiting the graph structures and semantic properties. 
Examples include 1) graph-level pretext task, which either distinguishes subgraphs of a certain node from those of other vertices~\cite{qiu2020gcc} or maximize the  mutual information between the local node representation and the global graph representations\cite{velickovic2019deep,infograph}. 2) Node-level task, which perform  node feature and edge generation~\cite{gpt_gnnhu} pretext tasks. 3) Hybrid-level task, which considers both node and graph-level tasks~\cite{you2020does,huiclr20gnnpretraining}. However, none of these models explore pre-training GNNs for recommendation, and we are the first to study the problem and define the reconstruction of the cold-start user/item embeddings as the pretext task.

\section{Conclusion}
This work explores pre-training a GNN model for addressing the cold-start recommendation problem. 
We propose a pretext task as reconstructing cold-start user/item embeddings to explicitly improve their embedding quality. We further incorporate a self-attention-based meta aggregator to improve the aggregation ability of each graph convolution step, and propose a sampling strategy to adaptively sample neighbors according to the GNN performance. 
Experiments on three datasets demonstrate the effectiveness of our proposed pre-training GNN against the original GNN models. 
We will explore multiple pretext tasks in the future work. 

\section*{ACKNOWLEDGMENTS}
This work is supported by National Key R\&D Program of China (No.2018YFB1004401), NSFC  (No.61532021, 61772537, 61772536, 61702522, 62076245), CCF-Tencent Open Fund and Australian Research Council (Grant No. DP190101985, DP170103954).

\clearpage
\balance
\normalem 
\bibliographystyle{ACM-Reference-Format}
\bibliography{sample-base}

\appendix

\end{document}